\documentclass[aps,prl,groupedaddress,superscriptaddress,twocolumn]{revtex4}

\usepackage{graphicx, amsmath, subfigure}
\usepackage{hyperref}
\usepackage{color}
\usepackage{float}
\usepackage{chngcntr}

\newcommand{\tabincell}[2]{\begin{tabular}{@{}#1@{}}#2\end{tabular}}

\begin{document}

\preprint{APS/123-QED}


\title{Non-Abelian braiding in spin superconductors utilizing the Aharonov-Casher effect}

\author{Yijia Wu}
\affiliation{International Center for Quantum Materials, School of Physics, Peking University, Beijing 100871, China}

\author{Jie Liu}
\affiliation{Department of Applied Physics, School of Science, Xian Jiaotong University, Xian 710049, China}

\author{Hua Jiang}
\affiliation{School of Physical Science and Technology, Soochow University, Suzhou 215006, China}

\author{Hua Chen}
\affiliation{Department of Physics, Zhejiang Normal University, Jinhua 321004, China}

\author{Haiwen Liu}
\affiliation{Center for Advanced Quantum Studies, Department of Physics, Beijing Normal University, Beijing 100875, China}

\author{X. C. Xie}
\thanks{Corresponding author: xcxie@pku.edu.cn}
\affiliation{International Center for Quantum Materials, School of Physics, Peking University, Beijing 100871, China}
\affiliation{Beijing Academy of Quantum Information Sciences, Beijing 100193, China}
\affiliation{CAS Center for Excellence in Topological Quantum Computation, University of Chinese Academy of Sciences, Beijing 100190, China}

\date{\today}

\begin{abstract}
Spin superconductor (SSC) is an exciton condensate state where the spin-triplet exciton superfluidity is charge neutral while spin $2(\hbar/2)$. In analogy to the Majorana zero mode (MZM) in topological superconductors, the interplay between SSC and band topology will also give rise to a specific kind of topological boundary state obeying non-Abelian braiding statistics. Remarkably, the non-Abelian geometric phase here originates from the Aharonov-Casher effect of the ``half-charge'' other than the Aharonov-Bohm effect. Such topological boundary state of SSC is bound with the vortex of electric flux gradient and can be experimentally more distinct than the MZM for being electrically charged. This theoretical proposal provides a new avenue investigating the non-Abelian braiding physics without the assistance of MZM and charge superconductor.
\end{abstract}

\maketitle



\textit{Introduction.} Stable excitons (electron-hole pairs) \cite{Keldysh&Kopaev, Kozlov&Maksimov, Keldysh&Kozlov, Lozovik&Yudson_sptatial_separated_exciton,
Shevchenko_sptatial_separated_exciton, Exciton_condensate_QSHI, RRDu_topological_exciton_condensate, LHHu_Exciton_condensate_Rashba, RRDu_CoulombDrag} usually emerge in a system where the electron is spatially well separated from the hole, or the electron's energy is lower than the hole's so that the electron-hole recombination is prohibited in low temperature. The exciton is surely charge neutral, nonetheless, it will possess non-vanishing electric dipole \cite{Lozovik&Yudson_sptatial_separated_exciton, Shevchenko_sptatial_separated_exciton, EDS_nature_Eisenstein&MacDonald, ExcitonCondensate_TIfilm, EDS_Eisenstein_review,  QDJiang_EDS} in the former case mentioned above, for instance, in the bilayer quantum Hall systems \cite{EDS_nature_Eisenstein&MacDonald, EDS_Eisenstein_review}. In the latter case, if the electron band is lower than the hole one due to the Zeeman splitting \cite{Sun&Xie_SSC_FerromagneticGraphene, Sun&Xie_SSC_Graphene0LL}, then the exciton is spin-polarized and possesses non-zero magnetic dipole moment.

Since the Coulomb interaction between electron and hole is naturally attractive, the mean field Hamiltonian of exciton condensate (EC) \cite{Sun&Xie_SSC_FerromagneticGraphene} resembles the Bogoliubov-de Gennes (BdG) Hamiltonian of the charge superconductor, where the Coulomb attraction plays the role of superconducting order parameter. A superconducting gap is thus formed in the EC so that the electron band and the hole band become mixed.
In such a way, a new ground state composed of excitons as ``Cooper pairs'' is formed and the EC can be named as  electric dipole superconductor (EDS) \cite{QDJiang_EDS} or spin superconductor (SSC) \cite{Sun&Xie_SSC_FerromagneticGraphene, Sun&Xie_SSC_Graphene0LL, ZQBao_SSC_GLtheory}, if the exciton here carries non-zero electric dipole or non-zero spin, respectively, as discussed above.

Both the EDS and the SSC share plenty of similarities with the conventional electric charge superconductor. For illustration, the charge neutral and spin $2(\hbar/2)$ ``Cooper pair'' in SSC is also a zero-momentum pair flowing dissipationlessly \cite{Sun&Xie_SSC_FerromagneticGraphene}. Additionally, the super spin current inside the SSC under an electromagnetic field can also be depicted by the London-type equations \cite{Sun&Xie_SSC_FerromagneticGraphene}. Meanwhile, the superconducting order parameter in SSC also satisfies the Ginzburg-Landau-type  theory \cite{ZQBao_SSC_GLtheory}.
Nevertheless, in contrast to its charge superconductor cousin that both the theories and the experiments of topological superconductor (TSC) have been studied intensively \cite{KitaevChain, Fu&Kane_proximity_MZM, MZM_semiconductor_SC_nanowire, MZM_signatures_Mourik_2012, MZM_JinFengJia, HongDing_ironMZM}, less attention has been paid to the interplay between  EC and band topology \cite{Vortex_in_EDS, ExcitonCondensation_InAsGaSb, TopologicalExcitonCondensate_TIfilm, ExcitonCondensate_TIfilm, RRDu_topological_exciton_condensate, HelicalTopological_ExcitonCondensate}, especially the possible fractional statistics \cite{Vortex_in_EDS} obeyed by the topological boundary states in EC.

\begin{table*}[t]
\begin{center}
\begin{tabular}{c c c c c c c c c}

\hline 
\hline
\rule{0pt}{6ex} 
\rule[-5.5ex]{0pt}{0pt} 
& \multicolumn{2}{c}{\tabincell{c}{Charge superconductor\\ with Hamiltonian \\ basis $(\begin{array}{cc}
c_{\mathbf{k}} & c_{-\mathbf{k}}^{\dagger}
\end{array} )^{\mathrm{T}}$} } 
& \quad \quad & \multicolumn{2}{c}{\tabincell{c}{Electric dipole superconductor \\ (EDS) with Hamiltonian \\ basis $( \begin{array}{cc}
c_{\mathbf{k}e,t} & c_{\mathbf{k}h,b}^{\dagger}
\end{array} )^{\mathrm{T}}$}} 
& \quad \quad & \multicolumn{2}{c}{\tabincell{c}{Spin superconductor (SSC) \\ with Hamiltonian \\ basis $(\begin{array}{cc}
c_{\mathbf{k}e\uparrow} & c_{\mathbf{k}h\uparrow}^{\dagger}
\end{array} )^{\mathrm{T}}$}}

\tabularnewline

\hline 
\rule{0pt}{4ex}
& \tabincell{c}{Cooper\\ pair} & \tabincell{c}{Majorana\\ zero mode} 
& & ``Cooper pair'' & \tabincell{c}{Topological\\ boundary state} 
& & ``Cooper pair'' & \tabincell{c}{Topological\\ boundary state} 
 
\tabularnewline

\rule{0pt}{3.5ex}
Expression 
& $c_{\mathbf{k}}^{\dagger}c_{-\mathbf{k}}^{\dagger}$ 
& $(c_{\mathbf{k}}\pm c_{-\mathbf{k}}^{\dagger})\big|_{\mathbf{k}=0}$ 
& & $c_{\mathbf{k}e,t}^{\dagger}c_{\mathbf{k}h,b}^{\dagger}$ 
& $(c_{\mathbf{k}e,t}\pm c_{\mathbf{k}h,b}^{\dagger})\big|_{\mathbf{k}=0}$ 
& & $c_{\mathbf{k}e\uparrow}^{\dagger}c_{\mathbf{k}h\uparrow}^{\dagger}$ 
& $(c_{\mathbf{k}e\uparrow}\pm c_{\mathbf{k}h\uparrow}^{\dagger})\big|_{\mathbf{k}=0}$

\tabularnewline

\rule{0pt}{3.5ex}  
Electric charge 
& $2|e|$ & $\left(-|e|\right)+|e|=0$ 
& & $0$ & $|e|$ 
& & $0$ & $|e|$

\tabularnewline

\rule{0pt}{3.5ex} 
Electric dipole & $0$ & $0$ 
& & $p_{0}$ & $0$ 
& & Undetermined & $0$

\tabularnewline

\rule{0pt}{3.5ex} 
Magnetic dipole & $\hbar$ & $\hbar/2$ 
& & Undetermined & Undetermined 
& & $\hbar$ & $\left(-\hbar/2\right)+\hbar/2=0$

\tabularnewline

\rule{0pt}{3.5ex}
\rule[-2ex]{0pt}{0pt} 
AB/HMW/AC phase & $2\pi$ & $\pi = \left( -\pi \ \mathrm{mod} \ 2\pi \right)$ 
& & $2\pi$ & $0$ 
& & $2\pi$ & $\pi = \left( -\pi \ \mathrm{mod} \ 2\pi \right)$

\tabularnewline

\hline
\hline  
\end{tabular}
\par\end{center}
\caption{Comparison for the bulk state (``Cooper pair'') and the topological boundary state of the charge superconductor, EDS, and SSC with non-trivial topology. The topological boundary state in SSC picks up a geometric phase of $\pi$ when it encircles a vortex, which is in parallel with the MZM in charge superconductor.}
\label{ComparisonTable}
\end{table*}

An important reason for TSC being appealing in the past decades is that its topological edge state, the Majorana zero mode (MZM), is regarded to obey the intriguing non-Abelian statistics and therefore as a promising candidate for topological quantum computation \cite{IvanovPRL2001, TQC_review_RMP}. In a two-dimensional TSC, the MZMs are bound with the magnetic vortices and form a degenerate ground state. Swapping any two vortices will give rise to a geometric phase of $\pi$ whose corresponding unitary transformation in the ground state Fock space is non-Abelian \cite{IvanovPRL2001}. Such a crucial $\pi$ phase is attributed to the periodic condition of the vortex as well as the ``half''-charge of the MZM. The periodicity of the wavefunction encircling the vortex implies that a Cooper pair $c_{\mathbf{k}}^{\dagger}c_{-\mathbf{k}}^{\dagger}$ inside the bulk of the TSC must pick up a $2\pi$ phase when it travels around an elementary magnetic vortex. The geometric phase $\phi = \oint \mathrm{d}\mathbf{r} \cdot (-q\mathbf{A})$ here is actually an Aharonov-Bohm (AB) phase \cite{AB_effect} where the charge of the Cooper pair is $q=-2|e|$. In comparison, the MZM $(c_{\mathbf{k}}\pm c_{-\mathbf{k}}^{\dagger})\big|_{\mathbf{k}=0}$ is an equally weighted superposition of an electron state and a hole state, where the charge of the electron and the hole composition is exactly half of the Cooper pair's as $-|e|$ and $|e|$, respectively. Hence, the AB phase accumulated by a MZM traveling around a vortex is $\pi$ by noticing that $\pi = -\pi \left(\mathrm{mod} \ 2\pi \right)$, although the MZM seems charge neutral.
In summary, the ``half''-charge property of the MZMs together with the periodicity and the ground state degeneracy lead to the non-Abelian braiding statistics.

In this Letter, we propose that such ``half-charge'' of the topological boundary state and its consequent non-Abelian braiding are also exhibited if a SSC is topologically non-trivial. The difference lies in that the ``charge'' in SSC refers spin other than the electric charge. Accordingly, the non-Abelian geometric phase here is an Aharonov-Casher (AC) phase \cite{AC_effect} other than an AB phase. 
The desired topological boundary state emerges and bounds with the vortex of electric flux gradient when the $p$-wave Coulomb attraction dominates in the SSC. Such type of vortex is quantized and could form the Abrikosov lattice due to the electric Meissner effect. Moreover, the subgap topological boundary state of SSC is electrically charged, hence it could be experimentally more distinct than the neutral MZM. 
This theory of spin-based non-Abelian braiding in the putative $p$-wave topological SSC opens a new way realizing the non-Abelian braiding beyond the conventional scheme utilizing the MZMs. In addition to the general theoretical discussion, we also put forward staggered ferromagnetic graphene as a possible material realizing the $p$-wave topological SSC.


\textit{Spin-based non-Abelian braiding utilizing the Aharonov-Casher effect.} From the relativistic theory of electrodynamics, the Lagrangian of a particle moving in the speed of $\mathbf{v}$ in an electromagnetic potential $\left( V, \mathbf{A} \right)$ is in the general form \cite{AC_effect} of $L = \frac{1}{2}m^{*}\mathbf{v}^{2} - [ q\left(V-\mathbf{v}\cdot\mathbf{A}\right)-\mathbf{p}_{0}\cdot\left(\mathbf{E}+\mathbf{v}\times\mathbf{B}\right)-\alpha\mathbf{s}\cdot\left(\mathbf{B}-\frac{\mathbf{v}\times\mathbf{E}}{c^{2}}\right) ]$, where $q$, $\mathbf{p}_0$, and $\mathbf{s}$ is the electric charge, the electric dipole, and the spin of this particle, respectively, $\mathbf{E}=-\nabla V$, and $\mathbf{B} = \nabla \times \mathbf{A}$ are the electromagnetic field. From this Lagrangian, a canonical momentum can be defined and the corresponding Peierls substitution has the form of

\begin{equation}
\mathbf{p} \to \mathbf{p} - q\mathbf{A} + \mathbf{p}_{0}\times\mathbf{B} - \alpha_{0}\mathbf{s}\times\mathbf{E}
\label{PeierlsSubstitution}
\end{equation}

\noindent where $\alpha_0 \equiv \alpha/c^2$ is the spin-orbit coupling (SOC) coefficient. The $-q\mathbf{A}$ term corresponds to the celebrated AB phase \cite{AB_effect} and leads to the non-Abelian statistics of the MZM as discussed above. In contrast, the He-McKellar-Wilkens (HMW) effect \cite{HMK_effect_HandM, HMK_effect_W, FluxQuantization_AC_and_HMW} from the $\mathbf{p}_{0}\times\mathbf{B}$ term and the AC effect from the $- \alpha_{0}\mathbf{s}\times\mathbf{E}$ term do not contribute when a MZM travels around a magnetic vortex, since a MZM does not carry any electric dipole $\mathbf{p}_0$ and a magnetic vortex does not contain an electric field $\mathbf{E}$.


Although being omitted in the usual discussions, the AC term $- \alpha_{0}\mathbf{s}\times\mathbf{E}$ plays an important role in the SSC. The Hamiltonian describing a two-dimensional SSC reads

\begin{equation}
H_{\mathrm{SSC}} = \sum_{\mathbf{k}}
\left( \begin{array}{cc} c_{\mathbf{k}e\uparrow}^{\dagger} & c_{\mathbf{k}h\uparrow}\end{array} \right)
\left( \begin{array}{cc}
\epsilon_{\mathbf{k}+} & \Delta_{\mathbf{k}} \\ \Delta_{\mathbf{k}}^{*} & \epsilon_{\mathbf{k}-}
\end{array} \right)
\left( \begin{array}{c} c_{\mathbf{k}e\uparrow} \\ c_{\mathbf{k}h\uparrow}^{\dagger} \end{array} \right)
\label{HamiltonianSSC}
\end{equation}

\noindent where $c_{\mathbf{k}e\uparrow}^{\dagger}$ creates an electron in the spin-up electron-like band, while $c_{\mathbf{k}h\uparrow}^{\dagger}$ creates a hole (annihilates an electron) in the spin-down hole-like band \cite{Sun&Xie_SSC_FerromagneticGraphene, Sun&Xie_SSC_Graphene0LL}. 
As a comparison, the Hamiltonian of a $p$-wave TSC can be expressed as $H_{\mathrm{TSC}} = \Delta(k_x\sigma_x - k_y\sigma_y) +(M-B|\mathbf{k}|^2)\sigma_z $ in the BdG basis $(c_{\mathbf{k}}, c_{-\mathbf{k}}^{\dagger})^{\mathrm{T}}$. Therefore, if both the dispersion of the electron-like band $\epsilon_{\mathbf{k}+}$ and the hole-like band $\epsilon_{\mathbf{k}-}$ in Eq. (\ref{HamiltonianSSC}) are quadratic and the $p$-wave pairing amplitude dominates in $\Delta_{\mathbf{k}}$, then such a SSC will also become topologically non-trivial.
By comparing the Hamiltonian basis of the SSC in Eq. (\ref{HamiltonianSSC}) and the BdG basis of the TSC, the topological boundary state of the SSC could be written as $(c_{\mathbf{k}e\uparrow}\pm c_{\mathbf{k}h\uparrow}^{\dagger})\big|_{\mathbf{k}=0}$, where $c_{\mathbf{k}e\uparrow}$ annihilates a spin-up electron in the electron-like band and $c_{\mathbf{k}h\uparrow}^{\dagger}$ creates a spin-up hole in the hole-like band.

For a topological SSC with a vortex in which the superconducting order parameter is suppressed, the periodic condition of the bulk state's wavefunction around the vortex requests that the ``Cooper pair'' state must acquire a $2\pi$ phase when it encircles this vortex. The ``Cooper pair'' here is a spin-polarized exciton $c_{\mathbf{k}e\uparrow}^{\dagger}c_{\mathbf{k}h\uparrow}^{\dagger}$ possessing spin $2\left(\hbar/2\right)\hat{z}$ so that such a $2\pi$ phase actually comes from the AC effect $\oint \mathrm{d}\mathbf{r} \cdot \left( -\alpha_{0}\mathbf{s}\times\mathbf{E} \right)$ \cite{AC_effect} [we will show later that the vortex in SSC does not contain a magnetic field, hence the other two terms in Eq. (\ref{PeierlsSubstitution}) do not contribute]. Therefore, in parallel to the MZM in TSC, although the topological boundary state of the SSC $(c_{\mathbf{k}e\uparrow}\pm c_{\mathbf{k}h\uparrow}^{\dagger})\big|_{\mathbf{k}=0}$ is seemingly spin zero, its two components with spin $\left(-\hbar/2\right) \hat{z}$ and $\left(\hbar/2\right) \hat{z}$ will both pick up an AC phase of $\pi$ by noticing again that $\pi = -\pi \left(\mathrm{mod} \ 2\pi \right)$. Such non-Abelian geometric phase presented in SSC are quite general as well as robust, since the only requirements here are the periodic condition and the ``half-charge'' carried by the subgap topological boundary state.

However, such ``half-charge''-related $\pi$ phase cannot be reached through the HMW effect in EDS, although its ``Cooper pair'' bulk state $c_{\mathbf{k}e,t}^{\dagger}c_{\mathbf{k}h,b}^{\dagger}$  (the subscript $t$ and $b$ denotes the top and the bottom layer in the bilayer system, respectively) with non-zero electric dipole $\mathbf{p}_0$ will also pick up a $2\pi$ phase as $\oint \mathrm{d}\mathbf{r} \cdot \left( \mathbf{p}_0\times\mathbf{B} \right)$ when it encircles a vortex in the EDS. This is because the topological boundary state $(c_{\mathbf{k}e,t}\pm c_{\mathbf{k}h,b}^{\dagger})\big|_{\mathbf{k}=0}$ here is a superposition of two single particle states, in which an individual single particle state cannot carry an electric dipole. Hence, in contrast to the charge superconductor and SSC, both the two components of the topological boundary state in EDS will gain a HMW phase of $0$ other than $\pm \pi$.


Now we have discussed the possibility to construct a non-Abelian geometric phase of $\pi$ utilizing all three terms in the Peierls substitution Eq. (\ref{PeierlsSubstitution}). The corresponding results are summarized in Table \ref{ComparisonTable}. It is worth noting that since the isolated magnetic monopole has thus far eluded discovery, such a spin (magnetic dipole)-based proposal is the minimal scheme to realize the non-Abelian braiding using the electromagnetic charge (monopole, dipole, quadrupole, etc.) besides the Majorana scheme utilizing the electric charge (electric monopole). Moreover, both the spin $\hbar/2$ in SSC and the electric charge $|e|$ in charge superconductor are elementary physics constants, thus the ``half-charge'' quantization and the resulting non-Abelian braiding properties are highly robust.

\begin{figure}[t]
\includegraphics[width=0.245\textwidth]{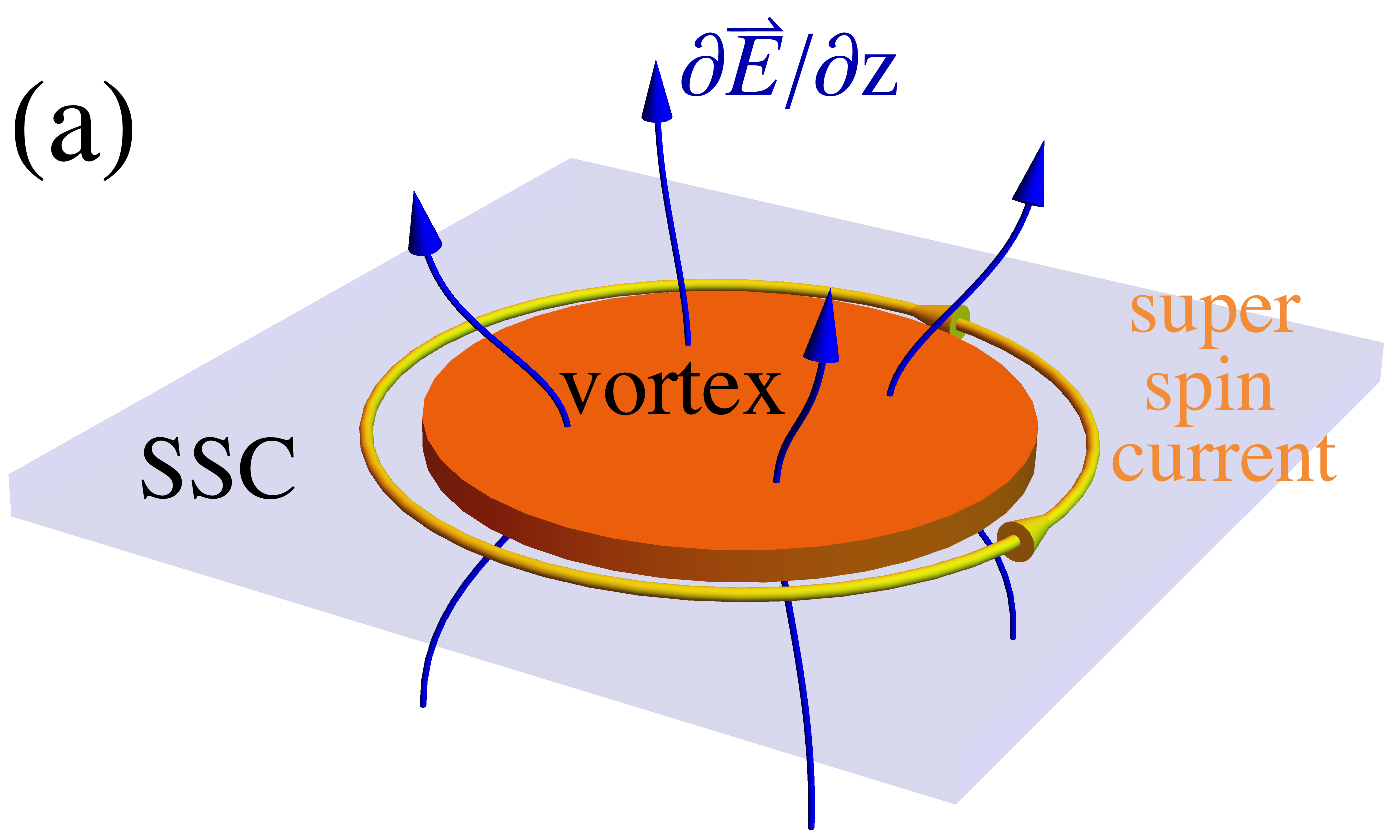}
\includegraphics[width=0.225\textwidth]{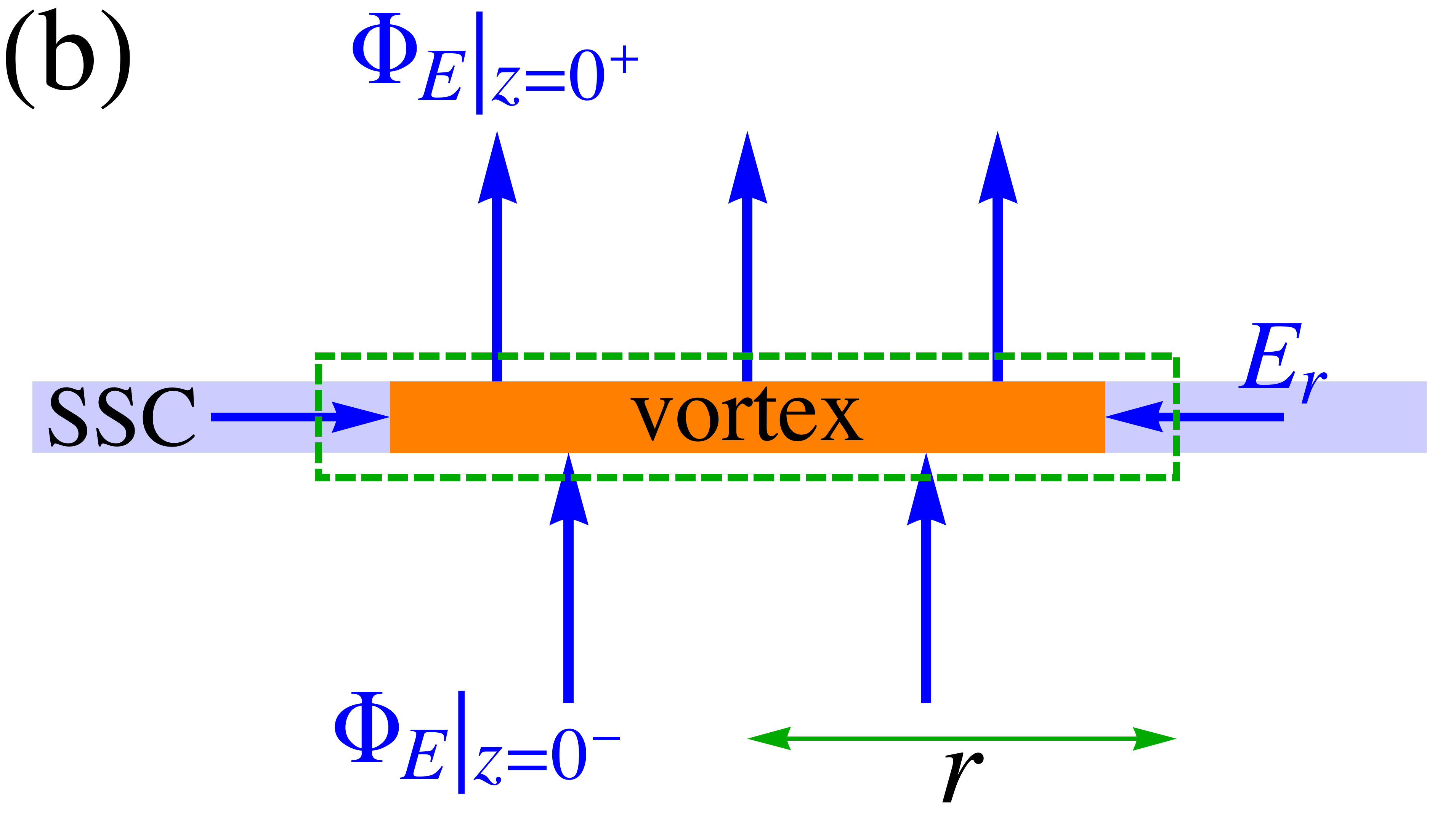}
\caption{(a) Sketch of a vortex in two-dimensional SSC that the gradient of the electric field along with the spin polarization direction $\partial \mathbf{E}/\partial z$ penetrates through the vortex while be screened out inside the SSC bulk. The circular yellow ring indicates the super spin current encircling the vortex.
(b) Side view of the Gaussian surface (indicated by the green dashed line) enclosing the vortex of the SSC.}
\label{sketch_vortex}
\end{figure}

\textit{Vortex of electric flux gradient.} As a consequence of the London-type equations of SSC \cite{Sun&Xie_SSC_FerromagneticGraphene}, the electric Meissner effect requests that the spatial variation of the electric field (other than the magnetic field) along with the spin polarization is screened out inside the SSC bulk. If we punch a hole in a two-dimensional SSC and impose an electric field with spatial gradient, then a circular super spin current will be induced around this hole and thus a vortex of electric flux gradient is formed (see Fig. \ref{sketch_vortex}(a), noticing that a pure spin current will not induce a magnetic field \cite{QFSun_SpinCurrentInducedField}). Since there is no net electric charge inside the vortex, the Gauss's law for electrostatics [see Fig. \ref{sketch_vortex}(b)] implies $\partial \Phi_E / \partial z = -2\pi r\cdot E_r$, where $\Phi_E$ is the electric flux penetrating the vortex, and $E_r$ is the electric field along the radial direction. Therefore the AC phase accumulated by moving a spin around such a vortex is

\begin{equation}
\oint \mathrm{d}\mathbf{r} \cdot \left( -\alpha_{0}\mathbf{s}\times\mathbf{E} \right) = 2\pi r \cdot \left( -\alpha_{0}sE_r \right) = -\alpha_0 s \frac{\partial \Phi_E}{\partial z}
\label{phase_electrix_flux_gradient}
\end{equation}

\noindent where the spin is assumed to be polarized along the $\hat{z}$-direction as $\mathbf{s}=s\hat{z}$. It follows that the electric flux gradient $\partial \Phi_E/\partial z$ in the SSC plays the role of the magnetic flux $\Phi_B$ in the TSC that $\oint \mathrm{d}\mathbf{r} \cdot \left( -q\mathbf{A} \right) = -q\Phi_B $.

Equation (\ref{phase_electrix_flux_gradient}) leads to the quantization of the electric flux gradient inside the vortex as $-\alpha_0 \hbar \left( \partial\Phi_E/\partial z\right) = 2\pi$, where the spin of the ``Cooper pair'' inside the SSC bulk $\mathbf{s} = \hbar \hat{z}$ has been inserted. Dealing with the $p$-wave topological SSC with such a vortex in the polar coordinate gives a zero-energy $E=0$ vortex-bounded state \cite{ShunQingShen_BoundState_inTI, YJWu_PRL, SupplementaryMaterials}

\begin{equation}
\psi_0 = \frac{C}{\sqrt{r}} \left[ e^{-\lambda_{+}\left(r-R\right)} - e^{-\lambda_{-}\left(r-R\right)} \right] \left( e^{i\theta}c_{\mathbf{k}h\uparrow}^{\dagger} + i c_{\mathbf{k}e\uparrow} \right) \Big|_{\mathbf{k}=0}
\label{ZeroMode}
\end{equation}

\noindent where $\lambda_{\pm}$ indicates the inverse of the localization length of this bounded zero mode, and $C$ is the normalization constant. Such an analytical solution confirms that the topological boundary state in the SSC is a zero mode carrying charge $|e|$ and spin $0$.

As in the case of charge superconductor, the vortices of the electric flux gradient in a SSC can also form periodic spatial units and thus constitute an Abrikosov flux lattice \cite{AbrikosovLattice1957, Abrikosov_Nobel_lecture, Superconductivity_Annett}. Assuming that an electric field with uniform gradient $\mathbf{E} = \left( -xE_0/L, 0, zE_0/L \right)$ is applied, then the linearized \cite{Superconductivity_Annett} first Ginzburg-Landau-type equation of the SSC \cite{ZQBao_SSC_GLtheory} gives rise to a Landau-level-like ground state solution for the bulk state as \cite{Superconductivity_Annett}

\begin{equation}
\Psi(\mathbf{r}) = \sum_{n_y} C_{0,n_{y}} \exp\left[-\frac{m^{*}\omega_{0}}{2\hbar} \left(x+\frac{2\pi L n_{y}}{\alpha_{0}E_{0}l_{y}}\right)^{2}\right] e^{i\frac{2\pi n_y}{l_y}y}
\label{GroundState_Abrikosov}
\end{equation}

\noindent where $l_y$ is the wave length along the $\hat{y}$-direction and $n_y$ is an integer. Here we have introduced the cyclotron frequency $ \omega_{0} = \frac{\alpha_{0}\hbar E_{0}}{m^* L}$ where $m^*$ is the effective mass of the exciton. As Abrikosov once assumed \cite{AbrikosovLattice1957, Abrikosov_Nobel_lecture}, if we assert that the ground state wavefunction $\Psi_0$ in Eq. (\ref{GroundState_Abrikosov}) is also periodic along the $\hat{x}$-direction with period $l_{x}$, then it requires the coefficients to be periodic as $C_{0,n_{y}+\nu}=C_{0,n_{y}}$ so that $l_x = \nu\frac{2\pi L}{\alpha_{0}E_{0}l_{y}}$ in which $\nu$ is an integer. Therefore, in each unit cell with area $l_x l_y$, the electric flux gradient penetrated is $\partial\Phi_E/\partial z = \frac{2\pi\nu}{\alpha_{0}}$, which is exactly the quantization condition of the electric flux gradient $-\alpha_0 \hbar \left( \partial\Phi_E/\partial z\right) = -\nu h$. Therefore, each unit cell here contains $v$ elementary vortices of electric flux gradient and the Abrikosov lattice is formed. The AC phase acquired by moving a vortex-bounded state $\psi_0$ [see Eq. (\ref{ZeroMode})] around a unit cell is $\oint \mathrm{d}\mathbf{r} \cdot \alpha_0 \frac{\hbar}{2} \frac{E_0}{L} x \hat{\mathbf{y}} = \alpha_0 \frac{\hbar}{2} \frac{E_0}{L} \left( -l_xl_y \right) = -\nu h/2$ as expected.


\textit{Discussions and possible material.} Such spin-based non-Abelian braiding originates from the ``half-charge'' property of the boundary state protected by the topology of the SSC, hence it is quite general and does not rely on the detailed parameters of the material. Here we propose the ferromagnetic graphene with staggered potential applied as a  possible candidate for $p$-wave topological SSC in which the single-particle bands $\epsilon_{\mathbf{k}\pm}$ are quadratic and the superconducting pairing potential $\Delta_{\mathbf{k}}$ is $p$-wave [see Eq. (\ref{HamiltonianSSC})]. 
First of all, the single-particle bands of the staggered ferromagnetic graphene are in the form of $E_{\mathbf{k}}^{\sigma,\pm} = \sigma M_F \pm \sqrt{t^2|\mathbf{k}|^2+V^2}$ in the vicinity of the Dirac points $\mathbf{K}, \mathbf{K}'$, where $\sigma = \uparrow,\downarrow$ represents the spin polarization, $M_F$ is the spin splitting energy, $t$ represents the nearest neighbor (NN) hopping amplitude in graphene, and the staggered potential $\pm V$ is applied to the different sublattices of the graphene. Therefore, the dispersions for the two lower bands can be approximately represented in a quadratic form as $\epsilon_{\mathbf{k}\pm} = \mp \left( M-B|\mathbf{k}|^2 \right)$. 
In addition, the pairing amplitude $\Delta_{\mathbf{k}}$ between the electron and the hole can be approximately taken into consideration by including the on-site Coulomb attraction and the NN Coulomb attraction in graphene as $\Delta_{\mathbf{k}} \propto U_{00}^{aa} + \frac{\sqrt{3}}{2} U_{0\delta}^{ab} \left( k_x \pm i k_y \right)$ \cite{SupplementaryMaterials}, where $\mathbf{k}$ is the wave vector near the Dirac points $\mathbf{K}, \mathbf{K}'$. Here, the $s$-wave pairing term is the on-site Coulomb attraction $U_{00}^{aa}$, while the $p$-wave term coming from the NN Coulomb attraction $U_{0\delta}^{ab}$ is protected by the lattice symmetry of the graphene \cite{graphene_pWave}. Thus, the pairing potential of the staggered ferromagnetic graphene is a mixture of the $s$-wave and the $p$-wave pairing. The $p$-wave pairing dominates and such a mixed-pairing SSC becomes topologically non-trivial in the condition that $U_{00}^{aa} < \frac{\sqrt{3}}{2} U_{0\delta}^{ab} \sqrt{M/B}$ \cite{SupplementaryMaterials}. Hence, although the on-site Coulomb attraction term is typically larger than the NN Coulomb attraction as $U_{00}^{aa} > U_{0\delta}^{ab}$, the staggered ferromagnetic graphene can still be a $p$-wave topological SSC if the flat band condition $\sqrt{M/B} \gg 1$ is satisfied. Remarkably, the zero energy of the topological boundary state will not shift up to the lowest order even when such a mixed pairing potential is presented \cite{SupplementaryMaterials}.

As we have discussed above and summarized in Table \ref{ComparisonTable}, the topological boundary state in the SSC is electrically charged and spin $0$, this is in stark contrast to the exciton bulk state in the SSC which is electrically neutral while possesses spin $2\left(\hbar/2\right)$. Taking the advantage of such a character, the subgap topological boundary state in SSC can hopefully be detected by the state-of-the-art spin-resolved probing technology \cite{SpinSelectiveAndreevReflection, SpinMZM_STM_Yazdani}. What is more, compared with the neutral MZM, the electrically-charged topological boundary state in the SSC can be much more distinct, for instance, in the experiments of resonant tunneling or Coulomb blockade \cite{Transport_QuantumDot, CoulombBlockade_Beenakker, ShouChengZhang_charge_pump}. 


\begin{figure}[t]
\includegraphics[width=0.35\textwidth]{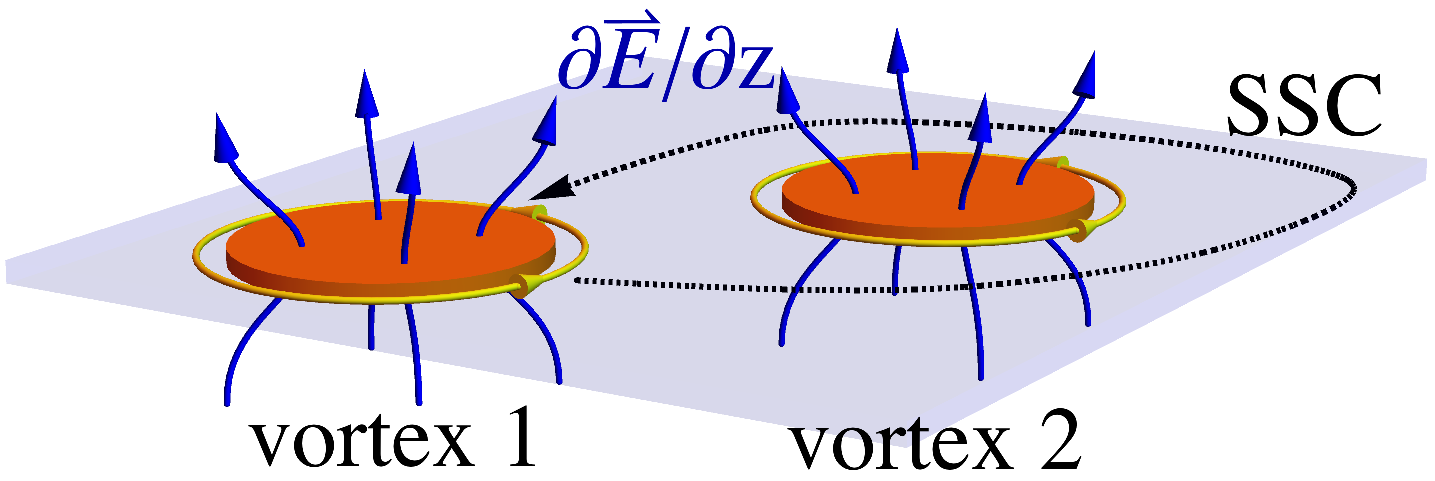}
\caption{Sketch of the braiding process that one vortex encircles another. The motion of the vortex-bounded topological boundary state will induce an undesired magnetic field.}
\label{braiding_sketch}
\end{figure}

However, the electric charge carried by the topological boundary state could also do harm to its non-Abelian braiding properties. 
Although the pure spin current \cite{QFSun_SpinCurrentInducedField} carried by the ``Cooper pair'' $c_{\mathbf{k}e\uparrow}^{\dagger}c_{\mathbf{k}h\uparrow}^{\dagger}$ creates no magnetic field, an undesirable magnetic field $\mathbf{A}$ will be induced if a charge current is also presented in the SSC. For instance, moving the topological boundary states during the braiding (see Fig. \ref{braiding_sketch})  will inevitably lead to a charge current. In such condition, an AB phase enters into the geometric phase accumulated by the electrically-charged topological boundary state as $\oint \mathrm{d}\mathbf{r} \cdot \left[ -\alpha_0 \left(\hbar/2\right) \hat{z} \times \mathbf{E} -|e|\mathbf{A} \right]$. Such a phase will deviate from $\pi$ since  the quantization condition of the charge neutral exciton $\oint \mathrm{d}\mathbf{r} \cdot \left( -\alpha_0\hbar\hat{z} \times \mathbf{E} \right) = 2\pi$ remains unchanged. Consequently, an additional adiabatic condition on the braiding time scale is required to reduce the charge current magnitude and thus eliminate such an undesired AB phase during the braiding.


Finally, it is worth noting that the topological boundary state in the SSC [cf. $\psi_0$ in Eq. (\ref{ZeroMode})] is actually a Dirac fermionic state other than a self-conjugate Majorana mode. Therefore, the quantum dimension spanned by each of such topological boundary state is $2$ other than $\sqrt{2}$, where the latter is the case for MZM since two independent MZMs could combine into a single Dirac fermionic mode. Thus, different from the Majorana case in which the braiding operations will not alter the fermion parity \cite{IvanovPRL2001}, the braiding operations of such topological boundary states conserve the fermion occupation number \cite{NuclearPhysicsB_DiracFermion, YJWu_PRL}. Correspondingly, the braiding matrices \cite{YJWu_PRL} as well as the fusion rules \cite{YJWu_NSR} of such topological boundary state also possess its own specific forms.


\textit{Summary.} We have proposed a theoretical framework of spin-based non-Abelian braiding in SSC utilizing the AC effect. The desired non-Abelian geometric phase of $\pi$ in the Fock space spanned by the topological boundary state only requires the periodicity of the exciton wavefunction around the vortex as well as the ``half-charge'' property of the subgap topological boundary state. Therefore, such proposal is quite universal since the periodicity is naturally satisfied, the ``half-charge'' $\hbar/2$ is a robust physics constant, and the subgap property is protected by the band topology. The possible experimental investigation requires a SSC material in which the $p$-wave pairing amplitude dominates and meanwhile, the possible disturbance from the AB effect could also be managed to rule out.


\textit{Acknowledgements.} We thank Qing-Feng Sun and Chui-Zhen Chen for fruitful discussion. This work is financially supported by the National Natural Science Foundation of China (Grants No. 11504008, No. 11534001, No. 11704338, No. 11822407, No. 11974271, and No. 12022407), the National Basic Research Program of China (Grants No. 2015CB921102 and No. 2019YFA0308403), and China Postdoctoral Science Foundation (Grant No. 2021M690233).

\bibliography{nonAbelian_SSC}

\providecommand{\noopsort}[1]{}\providecommand{\singleletter}[1]{#1}%
\begin{thebibliography}{48}
\expandafter\ifx\csname natexlab\endcsname\relax\def\natexlab#1{#1}\fi
\expandafter\ifx\csname bibnamefont\endcsname\relax
  \def\bibnamefont#1{#1}\fi
\expandafter\ifx\csname bibfnamefont\endcsname\relax
  \def\bibfnamefont#1{#1}\fi
\expandafter\ifx\csname citenamefont\endcsname\relax
  \def\citenamefont#1{#1}\fi
\expandafter\ifx\csname url\endcsname\relax
  \def\url#1{\texttt{#1}}\fi
\expandafter\ifx\csname urlprefix\endcsname\relax\def\urlprefix{URL }\fi
\providecommand{\bibinfo}[2]{#2}
\providecommand{\eprint}[2][]{\url{#2}}

\bibitem[{\citenamefont{Keldysh and Kopaev}(1965)}]{Keldysh&Kopaev}
\bibinfo{author}{\bibfnamefont{L.}~\bibnamefont{Keldysh}} \bibnamefont{and}
  \bibinfo{author}{\bibfnamefont{Y.~V.} \bibnamefont{Kopaev}},
  \bibinfo{journal}{Soviet Physics Solid State} \textbf{\bibinfo{volume}{6}},
  \bibinfo{pages}{2219} (\bibinfo{year}{1965}).

\bibitem[{\citenamefont{Kozlov and Maksimov}(1965)}]{Kozlov&Maksimov}
\bibinfo{author}{\bibfnamefont{A.}~\bibnamefont{Kozlov}} \bibnamefont{and}
  \bibinfo{author}{\bibfnamefont{L.}~\bibnamefont{Maksimov}},
  \bibinfo{journal}{Sov. Phys. JETP} \textbf{\bibinfo{volume}{21}},
  \bibinfo{pages}{790} (\bibinfo{year}{1965}).

\bibitem[{\citenamefont{Keldysh and Kozlov}(1968)}]{Keldysh&Kozlov}
\bibinfo{author}{\bibfnamefont{L.}~\bibnamefont{Keldysh}} \bibnamefont{and}
  \bibinfo{author}{\bibfnamefont{A.}~\bibnamefont{Kozlov}},
  \bibinfo{journal}{Sov. Phys. JETP} \textbf{\bibinfo{volume}{27}},
  \bibinfo{pages}{521} (\bibinfo{year}{1968}).

\bibitem[{\citenamefont{Lozovik and
  Yudson}(1976)}]{Lozovik&Yudson_sptatial_separated_exciton}
\bibinfo{author}{\bibfnamefont{Y.}~\bibnamefont{Lozovik}} \bibnamefont{and}
  \bibinfo{author}{\bibfnamefont{V.}~\bibnamefont{Yudson}},
  \bibinfo{journal}{Sov. Phys. JETP} \textbf{\bibinfo{volume}{44}},
  \bibinfo{pages}{389} (\bibinfo{year}{1976}).

\bibitem[{\citenamefont{Shevchenko}(1976)}]{Shevchenko_sptatial_separated_exciton}
\bibinfo{author}{\bibfnamefont{S.}~\bibnamefont{Shevchenko}},
  \bibinfo{journal}{Sov. J. Low Temp. Phys.} \textbf{\bibinfo{volume}{2}},
  \bibinfo{pages}{251} (\bibinfo{year}{1976}).

\bibitem[{\citenamefont{Pikulin and
  Hyart}(2014{\natexlab{a}})}]{Exciton_condensate_QSHI}
\bibinfo{author}{\bibfnamefont{D.~I.} \bibnamefont{Pikulin}} \bibnamefont{and}
  \bibinfo{author}{\bibfnamefont{T.}~\bibnamefont{Hyart}},
  \bibinfo{journal}{Phys. Rev. Lett.} \textbf{\bibinfo{volume}{112}},
  \bibinfo{pages}{176403} (\bibinfo{year}{2014}{\natexlab{a}}).

\bibitem[{\citenamefont{Du et~al.}(2017)\citenamefont{Du, Li, Lou, Sullivan,
  Chang, Kono, and Du}}]{RRDu_topological_exciton_condensate}
\bibinfo{author}{\bibfnamefont{L.}~\bibnamefont{Du}},
  \bibinfo{author}{\bibfnamefont{X.}~\bibnamefont{Li}},
  \bibinfo{author}{\bibfnamefont{W.}~\bibnamefont{Lou}},
  \bibinfo{author}{\bibfnamefont{G.}~\bibnamefont{Sullivan}},
  \bibinfo{author}{\bibfnamefont{K.}~\bibnamefont{Chang}},
  \bibinfo{author}{\bibfnamefont{J.}~\bibnamefont{Kono}}, \bibnamefont{and}
  \bibinfo{author}{\bibfnamefont{R.-R.} \bibnamefont{Du}},
  \bibinfo{journal}{Nature communications} \textbf{\bibinfo{volume}{8}},
  \bibinfo{pages}{1} (\bibinfo{year}{2017}).

\bibitem[{\citenamefont{Hu et~al.}(2020)\citenamefont{Hu, Zhang, Zhang, and
  Wu}}]{LHHu_Exciton_condensate_Rashba}
\bibinfo{author}{\bibfnamefont{L.-H.} \bibnamefont{Hu}},
  \bibinfo{author}{\bibfnamefont{R.-X.} \bibnamefont{Zhang}},
  \bibinfo{author}{\bibfnamefont{F.-C.} \bibnamefont{Zhang}}, \bibnamefont{and}
  \bibinfo{author}{\bibfnamefont{C.}~\bibnamefont{Wu}}, \bibinfo{journal}{Phys.
  Rev. B} \textbf{\bibinfo{volume}{102}}, \bibinfo{pages}{235115}
  (\bibinfo{year}{2020}).

\bibitem[{\citenamefont{Du et~al.}(2021)\citenamefont{Du, Zheng, Chou, Zhang,
  Wu, Sullivan, Ikhlassi, and Du}}]{RRDu_CoulombDrag}
\bibinfo{author}{\bibfnamefont{L.}~\bibnamefont{Du}},
  \bibinfo{author}{\bibfnamefont{J.}~\bibnamefont{Zheng}},
  \bibinfo{author}{\bibfnamefont{Y.-Z.} \bibnamefont{Chou}},
  \bibinfo{author}{\bibfnamefont{J.}~\bibnamefont{Zhang}},
  \bibinfo{author}{\bibfnamefont{X.}~\bibnamefont{Wu}},
  \bibinfo{author}{\bibfnamefont{G.}~\bibnamefont{Sullivan}},
  \bibinfo{author}{\bibfnamefont{A.}~\bibnamefont{Ikhlassi}}, \bibnamefont{and}
  \bibinfo{author}{\bibfnamefont{R.-R.} \bibnamefont{Du}},
  \bibinfo{journal}{Nature Electronics} pp. \bibinfo{pages}{1--6}
  (\bibinfo{year}{2021}).

\bibitem[{\citenamefont{Eisenstein and
  MacDonald}(2004)}]{EDS_nature_Eisenstein&MacDonald}
\bibinfo{author}{\bibfnamefont{J.}~\bibnamefont{Eisenstein}} \bibnamefont{and}
  \bibinfo{author}{\bibfnamefont{A.}~\bibnamefont{MacDonald}},
  \bibinfo{journal}{Nature} \textbf{\bibinfo{volume}{432}},
  \bibinfo{pages}{691} (\bibinfo{year}{2004}).

\bibitem[{\citenamefont{Seradjeh et~al.}(2009)\citenamefont{Seradjeh, Moore,
  and Franz}}]{ExcitonCondensate_TIfilm}
\bibinfo{author}{\bibfnamefont{B.}~\bibnamefont{Seradjeh}},
  \bibinfo{author}{\bibfnamefont{J.~E.} \bibnamefont{Moore}}, \bibnamefont{and}
  \bibinfo{author}{\bibfnamefont{M.}~\bibnamefont{Franz}},
  \bibinfo{journal}{Phys. Rev. Lett.} \textbf{\bibinfo{volume}{103}},
  \bibinfo{pages}{066402} (\bibinfo{year}{2009}).

\bibitem[{\citenamefont{Eisenstein}(2014)}]{EDS_Eisenstein_review}
\bibinfo{author}{\bibfnamefont{J.}~\bibnamefont{Eisenstein}},
  \bibinfo{journal}{Annu. Rev. Condens. Matter Phys.}
  \textbf{\bibinfo{volume}{5}}, \bibinfo{pages}{159} (\bibinfo{year}{2014}).

\bibitem[{\citenamefont{Jiang et~al.}(2015)\citenamefont{Jiang, Bao, Sun, and
  Xie}}]{QDJiang_EDS}
\bibinfo{author}{\bibfnamefont{Q.-D.} \bibnamefont{Jiang}},
  \bibinfo{author}{\bibfnamefont{Z.-q.} \bibnamefont{Bao}},
  \bibinfo{author}{\bibfnamefont{Q.-F.} \bibnamefont{Sun}}, \bibnamefont{and}
  \bibinfo{author}{\bibfnamefont{X.~C.} \bibnamefont{Xie}},
  \bibinfo{journal}{Scientific reports} \textbf{\bibinfo{volume}{5}},
  \bibinfo{pages}{1} (\bibinfo{year}{2015}).

\bibitem[{\citenamefont{Sun et~al.}(2011)\citenamefont{Sun, Jiang, Yu, and
  Xie}}]{Sun&Xie_SSC_FerromagneticGraphene}
\bibinfo{author}{\bibfnamefont{Q.-f.} \bibnamefont{Sun}},
  \bibinfo{author}{\bibfnamefont{Z.-t.} \bibnamefont{Jiang}},
  \bibinfo{author}{\bibfnamefont{Y.}~\bibnamefont{Yu}}, \bibnamefont{and}
  \bibinfo{author}{\bibfnamefont{X.~C.} \bibnamefont{Xie}},
  \bibinfo{journal}{Phys. Rev. B} \textbf{\bibinfo{volume}{84}},
  \bibinfo{pages}{214501} (\bibinfo{year}{2011}).

\bibitem[{\citenamefont{Sun and Xie}(2013)}]{Sun&Xie_SSC_Graphene0LL}
\bibinfo{author}{\bibfnamefont{Q.-f.} \bibnamefont{Sun}} \bibnamefont{and}
  \bibinfo{author}{\bibfnamefont{X.~C.} \bibnamefont{Xie}},
  \bibinfo{journal}{Phys. Rev. B} \textbf{\bibinfo{volume}{87}},
  \bibinfo{pages}{245427} (\bibinfo{year}{2013}).

\bibitem[{\citenamefont{Bao et~al.}(2013)\citenamefont{Bao, Xie, and
  Sun}}]{ZQBao_SSC_GLtheory}
\bibinfo{author}{\bibfnamefont{Z.-q.} \bibnamefont{Bao}},
  \bibinfo{author}{\bibfnamefont{X.~C.} \bibnamefont{Xie}}, \bibnamefont{and}
  \bibinfo{author}{\bibfnamefont{Q.-f.} \bibnamefont{Sun}},
  \bibinfo{journal}{Nature communications} \textbf{\bibinfo{volume}{4}},
  \bibinfo{pages}{1} (\bibinfo{year}{2013}).

\bibitem[{\citenamefont{Kitaev}(2001)}]{KitaevChain}
\bibinfo{author}{\bibfnamefont{A.~Y.} \bibnamefont{Kitaev}},
  \bibinfo{journal}{Physics-uspekhi} \textbf{\bibinfo{volume}{44}},
  \bibinfo{pages}{131} (\bibinfo{year}{2001}).

\bibitem[{\citenamefont{Fu and Kane}(2008)}]{Fu&Kane_proximity_MZM}
\bibinfo{author}{\bibfnamefont{L.}~\bibnamefont{Fu}} \bibnamefont{and}
  \bibinfo{author}{\bibfnamefont{C.~L.} \bibnamefont{Kane}},
  \bibinfo{journal}{Phys. Rev. Lett.} \textbf{\bibinfo{volume}{100}},
  \bibinfo{pages}{096407} (\bibinfo{year}{2008}).

\bibitem[{\citenamefont{Lutchyn et~al.}(2010)\citenamefont{Lutchyn, Sau, and
  Das~Sarma}}]{MZM_semiconductor_SC_nanowire}
\bibinfo{author}{\bibfnamefont{R.~M.} \bibnamefont{Lutchyn}},
  \bibinfo{author}{\bibfnamefont{J.~D.} \bibnamefont{Sau}}, \bibnamefont{and}
  \bibinfo{author}{\bibfnamefont{S.}~\bibnamefont{Das~Sarma}},
  \bibinfo{journal}{Phys. Rev. Lett.} \textbf{\bibinfo{volume}{105}},
  \bibinfo{pages}{077001} (\bibinfo{year}{2010}).

\bibitem[{\citenamefont{Mourik et~al.}(2012)\citenamefont{Mourik, Zuo, Frolov,
  Plissard, Bakkers, and Kouwenhoven}}]{MZM_signatures_Mourik_2012}
\bibinfo{author}{\bibfnamefont{V.}~\bibnamefont{Mourik}},
  \bibinfo{author}{\bibfnamefont{K.}~\bibnamefont{Zuo}},
  \bibinfo{author}{\bibfnamefont{S.~M.} \bibnamefont{Frolov}},
  \bibinfo{author}{\bibfnamefont{S.}~\bibnamefont{Plissard}},
  \bibinfo{author}{\bibfnamefont{E.~P.} \bibnamefont{Bakkers}},
  \bibnamefont{and} \bibinfo{author}{\bibfnamefont{L.~P.}
  \bibnamefont{Kouwenhoven}}, \bibinfo{journal}{Science}
  \textbf{\bibinfo{volume}{336}}, \bibinfo{pages}{1003} (\bibinfo{year}{2012}).

\bibitem[{\citenamefont{Sun et~al.}(2016{\natexlab{a}})\citenamefont{Sun,
  Zhang, Hu, Li, Wang, Ma, Xu, Gao, Guan, Li et~al.}}]{MZM_JinFengJia}
\bibinfo{author}{\bibfnamefont{H.-H.} \bibnamefont{Sun}},
  \bibinfo{author}{\bibfnamefont{K.-W.} \bibnamefont{Zhang}},
  \bibinfo{author}{\bibfnamefont{L.-H.} \bibnamefont{Hu}},
  \bibinfo{author}{\bibfnamefont{C.}~\bibnamefont{Li}},
  \bibinfo{author}{\bibfnamefont{G.-Y.} \bibnamefont{Wang}},
  \bibinfo{author}{\bibfnamefont{H.-Y.} \bibnamefont{Ma}},
  \bibinfo{author}{\bibfnamefont{Z.-A.} \bibnamefont{Xu}},
  \bibinfo{author}{\bibfnamefont{C.-L.} \bibnamefont{Gao}},
  \bibinfo{author}{\bibfnamefont{D.-D.} \bibnamefont{Guan}},
  \bibinfo{author}{\bibfnamefont{Y.-Y.} \bibnamefont{Li}},
  \bibnamefont{et~al.}, \bibinfo{journal}{Phys. Rev. Lett.}
  \textbf{\bibinfo{volume}{116}}, \bibinfo{pages}{257003}
  (\bibinfo{year}{2016}{\natexlab{a}}).

\bibitem[{\citenamefont{Zhu et~al.}(2020)\citenamefont{Zhu, Kong, Cao, Chen,
  Papaj, Du, Xing, Liu, Wang, Shen et~al.}}]{HongDing_ironMZM}
\bibinfo{author}{\bibfnamefont{S.}~\bibnamefont{Zhu}},
  \bibinfo{author}{\bibfnamefont{L.}~\bibnamefont{Kong}},
  \bibinfo{author}{\bibfnamefont{L.}~\bibnamefont{Cao}},
  \bibinfo{author}{\bibfnamefont{H.}~\bibnamefont{Chen}},
  \bibinfo{author}{\bibfnamefont{M.}~\bibnamefont{Papaj}},
  \bibinfo{author}{\bibfnamefont{S.}~\bibnamefont{Du}},
  \bibinfo{author}{\bibfnamefont{Y.}~\bibnamefont{Xing}},
  \bibinfo{author}{\bibfnamefont{W.}~\bibnamefont{Liu}},
  \bibinfo{author}{\bibfnamefont{D.}~\bibnamefont{Wang}},
  \bibinfo{author}{\bibfnamefont{C.}~\bibnamefont{Shen}}, \bibnamefont{et~al.},
  \bibinfo{journal}{Science} \textbf{\bibinfo{volume}{367}},
  \bibinfo{pages}{189} (\bibinfo{year}{2020}).

\bibitem[{\citenamefont{Seradjeh et~al.}(2008)\citenamefont{Seradjeh, Weber,
  and Franz}}]{Vortex_in_EDS}
\bibinfo{author}{\bibfnamefont{B.}~\bibnamefont{Seradjeh}},
  \bibinfo{author}{\bibfnamefont{H.}~\bibnamefont{Weber}}, \bibnamefont{and}
  \bibinfo{author}{\bibfnamefont{M.}~\bibnamefont{Franz}},
  \bibinfo{journal}{Phys. Rev. Lett.} \textbf{\bibinfo{volume}{101}},
  \bibinfo{pages}{246404} (\bibinfo{year}{2008}).

\bibitem[{\citenamefont{Pikulin and
  Hyart}(2014{\natexlab{b}})}]{ExcitonCondensation_InAsGaSb}
\bibinfo{author}{\bibfnamefont{D.~I.} \bibnamefont{Pikulin}} \bibnamefont{and}
  \bibinfo{author}{\bibfnamefont{T.}~\bibnamefont{Hyart}},
  \bibinfo{journal}{Phys. Rev. Lett.} \textbf{\bibinfo{volume}{112}},
  \bibinfo{pages}{176403} (\bibinfo{year}{2014}{\natexlab{b}}).

\bibitem[{\citenamefont{Seradjeh}(2012)}]{TopologicalExcitonCondensate_TIfilm}
\bibinfo{author}{\bibfnamefont{B.}~\bibnamefont{Seradjeh}},
  \bibinfo{journal}{Phys. Rev. B} \textbf{\bibinfo{volume}{85}},
  \bibinfo{pages}{235146} (\bibinfo{year}{2012}).

\bibitem[{\citenamefont{Budich et~al.}(2014)\citenamefont{Budich, Trauzettel,
  and Michetti}}]{HelicalTopological_ExcitonCondensate}
\bibinfo{author}{\bibfnamefont{J.~C.} \bibnamefont{Budich}},
  \bibinfo{author}{\bibfnamefont{B.}~\bibnamefont{Trauzettel}},
  \bibnamefont{and} \bibinfo{author}{\bibfnamefont{P.}~\bibnamefont{Michetti}},
  \bibinfo{journal}{Phys. Rev. Lett.} \textbf{\bibinfo{volume}{112}},
  \bibinfo{pages}{146405} (\bibinfo{year}{2014}).

\bibitem[{\citenamefont{Ivanov}(2001)}]{IvanovPRL2001}
\bibinfo{author}{\bibfnamefont{D.~A.} \bibnamefont{Ivanov}},
  \bibinfo{journal}{Phys. Rev. Lett.} \textbf{\bibinfo{volume}{86}},
  \bibinfo{pages}{268} (\bibinfo{year}{2001}).

\bibitem[{\citenamefont{Nayak et~al.}(2008)\citenamefont{Nayak, Simon, Stern,
  Freedman, and Das~Sarma}}]{TQC_review_RMP}
\bibinfo{author}{\bibfnamefont{C.}~\bibnamefont{Nayak}},
  \bibinfo{author}{\bibfnamefont{S.~H.} \bibnamefont{Simon}},
  \bibinfo{author}{\bibfnamefont{A.}~\bibnamefont{Stern}},
  \bibinfo{author}{\bibfnamefont{M.}~\bibnamefont{Freedman}}, \bibnamefont{and}
  \bibinfo{author}{\bibfnamefont{S.}~\bibnamefont{Das~Sarma}},
  \bibinfo{journal}{Rev. Mod. Phys.} \textbf{\bibinfo{volume}{80}},
  \bibinfo{pages}{1083} (\bibinfo{year}{2008}).

\bibitem[{\citenamefont{Aharonov and Bohm}(1959)}]{AB_effect}
\bibinfo{author}{\bibfnamefont{Y.}~\bibnamefont{Aharonov}} \bibnamefont{and}
  \bibinfo{author}{\bibfnamefont{D.}~\bibnamefont{Bohm}},
  \bibinfo{journal}{Phys. Rev.} \textbf{\bibinfo{volume}{115}},
  \bibinfo{pages}{485} (\bibinfo{year}{1959}).

\bibitem[{\citenamefont{Aharonov and Casher}(1984)}]{AC_effect}
\bibinfo{author}{\bibfnamefont{Y.}~\bibnamefont{Aharonov}} \bibnamefont{and}
  \bibinfo{author}{\bibfnamefont{A.}~\bibnamefont{Casher}},
  \bibinfo{journal}{Phys. Rev. Lett.} \textbf{\bibinfo{volume}{53}},
  \bibinfo{pages}{319} (\bibinfo{year}{1984}).

\bibitem[{\citenamefont{He and McKellar}(1993)}]{HMK_effect_HandM}
\bibinfo{author}{\bibfnamefont{X.-G.} \bibnamefont{He}} \bibnamefont{and}
  \bibinfo{author}{\bibfnamefont{B.~H.~J.} \bibnamefont{McKellar}},
  \bibinfo{journal}{Phys. Rev. A} \textbf{\bibinfo{volume}{47}},
  \bibinfo{pages}{3424} (\bibinfo{year}{1993}).

\bibitem[{\citenamefont{Wilkens}(1994)}]{HMK_effect_W}
\bibinfo{author}{\bibfnamefont{M.}~\bibnamefont{Wilkens}},
  \bibinfo{journal}{Phys. Rev. Lett.} \textbf{\bibinfo{volume}{72}},
  \bibinfo{pages}{5} (\bibinfo{year}{1994}).

\bibitem[{\citenamefont{Chen et~al.}(2013)\citenamefont{Chen, Horsch, and
  Manske}}]{FluxQuantization_AC_and_HMW}
\bibinfo{author}{\bibfnamefont{W.}~\bibnamefont{Chen}},
  \bibinfo{author}{\bibfnamefont{P.}~\bibnamefont{Horsch}}, \bibnamefont{and}
  \bibinfo{author}{\bibfnamefont{D.}~\bibnamefont{Manske}},
  \bibinfo{journal}{Phys. Rev. B} \textbf{\bibinfo{volume}{87}},
  \bibinfo{pages}{214502} (\bibinfo{year}{2013}).

\bibitem[{\citenamefont{Sun et~al.}(2004)\citenamefont{Sun, Guo, and
  Wang}}]{QFSun_SpinCurrentInducedField}
\bibinfo{author}{\bibfnamefont{Q.-f.} \bibnamefont{Sun}},
  \bibinfo{author}{\bibfnamefont{H.}~\bibnamefont{Guo}}, \bibnamefont{and}
  \bibinfo{author}{\bibfnamefont{J.}~\bibnamefont{Wang}},
  \bibinfo{journal}{Phys. Rev. B} \textbf{\bibinfo{volume}{69}},
  \bibinfo{pages}{054409} (\bibinfo{year}{2004}).

\bibitem[{\citenamefont{Shan et~al.}(2011)\citenamefont{Shan, Lu, Lu, and
  Shen}}]{ShunQingShen_BoundState_inTI}
\bibinfo{author}{\bibfnamefont{W.-Y.} \bibnamefont{Shan}},
  \bibinfo{author}{\bibfnamefont{J.}~\bibnamefont{Lu}},
  \bibinfo{author}{\bibfnamefont{H.-Z.} \bibnamefont{Lu}}, \bibnamefont{and}
  \bibinfo{author}{\bibfnamefont{S.-Q.} \bibnamefont{Shen}},
  \bibinfo{journal}{Phys. Rev. B} \textbf{\bibinfo{volume}{84}},
  \bibinfo{pages}{035307} (\bibinfo{year}{2011}).

\bibitem[{\citenamefont{Wu et~al.}(2020{\natexlab{a}})\citenamefont{Wu, Jiang,
  Liu, Liu, and Xie}}]{YJWu_PRL}
\bibinfo{author}{\bibfnamefont{Y.}~\bibnamefont{Wu}},
  \bibinfo{author}{\bibfnamefont{H.}~\bibnamefont{Jiang}},
  \bibinfo{author}{\bibfnamefont{J.}~\bibnamefont{Liu}},
  \bibinfo{author}{\bibfnamefont{H.}~\bibnamefont{Liu}}, \bibnamefont{and}
  \bibinfo{author}{\bibfnamefont{X.~C.} \bibnamefont{Xie}},
  \bibinfo{journal}{Phys. Rev. Lett.} \textbf{\bibinfo{volume}{125}},
  \bibinfo{pages}{036801} (\bibinfo{year}{2020}{\natexlab{a}}).

\bibitem[{Sup()}]{SupplementaryMaterials}
\bibinfo{note}{See Supplementary Materials for the analytical solution of the
  vortex-bounded topological boundary state in the spin superconductor, and the
  Coulomb attraction in the staggered ferromagnetic graphene.}

\bibitem[{\citenamefont{Abrikosov}(1957)}]{AbrikosovLattice1957}
\bibinfo{author}{\bibfnamefont{A.~A.} \bibnamefont{Abrikosov}},
  \bibinfo{journal}{Sov. Phys. JETP} \textbf{\bibinfo{volume}{5}},
  \bibinfo{pages}{1174} (\bibinfo{year}{1957}).

\bibitem[{\citenamefont{Abrikosov}(2004)}]{Abrikosov_Nobel_lecture}
\bibinfo{author}{\bibfnamefont{A.~A.} \bibnamefont{Abrikosov}},
  \bibinfo{journal}{Rev. Mod. Phys.} \textbf{\bibinfo{volume}{76}},
  \bibinfo{pages}{975} (\bibinfo{year}{2004}).

\bibitem[{\citenamefont{Annett et~al.}(2004)}]{Superconductivity_Annett}
\bibinfo{author}{\bibfnamefont{J.~F.} \bibnamefont{Annett}}
  \bibnamefont{et~al.}, \emph{\bibinfo{title}{Superconductivity, superfluids
  and condensates}}, vol.~\bibinfo{volume}{5} (\bibinfo{publisher}{Oxford
  University Press}, \bibinfo{year}{2004}).

\bibitem[{\citenamefont{Uchoa and Castro~Neto}(2007)}]{graphene_pWave}
\bibinfo{author}{\bibfnamefont{B.}~\bibnamefont{Uchoa}} \bibnamefont{and}
  \bibinfo{author}{\bibfnamefont{A.~H.} \bibnamefont{Castro~Neto}},
  \bibinfo{journal}{Phys. Rev. Lett.} \textbf{\bibinfo{volume}{98}},
  \bibinfo{pages}{146801} (\bibinfo{year}{2007}).

\bibitem[{\citenamefont{Sun et~al.}(2016{\natexlab{b}})\citenamefont{Sun,
  Zhang, Hu, Li, Wang, Ma, Xu, Gao, Guan, Li
  et~al.}}]{SpinSelectiveAndreevReflection}
\bibinfo{author}{\bibfnamefont{H.-H.} \bibnamefont{Sun}},
  \bibinfo{author}{\bibfnamefont{K.-W.} \bibnamefont{Zhang}},
  \bibinfo{author}{\bibfnamefont{L.-H.} \bibnamefont{Hu}},
  \bibinfo{author}{\bibfnamefont{C.}~\bibnamefont{Li}},
  \bibinfo{author}{\bibfnamefont{G.-Y.} \bibnamefont{Wang}},
  \bibinfo{author}{\bibfnamefont{H.-Y.} \bibnamefont{Ma}},
  \bibinfo{author}{\bibfnamefont{Z.-A.} \bibnamefont{Xu}},
  \bibinfo{author}{\bibfnamefont{C.-L.} \bibnamefont{Gao}},
  \bibinfo{author}{\bibfnamefont{D.-D.} \bibnamefont{Guan}},
  \bibinfo{author}{\bibfnamefont{Y.-Y.} \bibnamefont{Li}},
  \bibnamefont{et~al.}, \bibinfo{journal}{Phys. Rev. Lett.}
  \textbf{\bibinfo{volume}{116}}, \bibinfo{pages}{257003}
  (\bibinfo{year}{2016}{\natexlab{b}}).

\bibitem[{\citenamefont{Jeon et~al.}(2017)\citenamefont{Jeon, Xie, Li, Wang,
  Bernevig, and Yazdani}}]{SpinMZM_STM_Yazdani}
\bibinfo{author}{\bibfnamefont{S.}~\bibnamefont{Jeon}},
  \bibinfo{author}{\bibfnamefont{Y.}~\bibnamefont{Xie}},
  \bibinfo{author}{\bibfnamefont{J.}~\bibnamefont{Li}},
  \bibinfo{author}{\bibfnamefont{Z.}~\bibnamefont{Wang}},
  \bibinfo{author}{\bibfnamefont{B.~A.} \bibnamefont{Bernevig}},
  \bibnamefont{and} \bibinfo{author}{\bibfnamefont{A.}~\bibnamefont{Yazdani}},
  \bibinfo{journal}{Science} \textbf{\bibinfo{volume}{358}},
  \bibinfo{pages}{772} (\bibinfo{year}{2017}).

\bibitem[{\citenamefont{Meir et~al.}(1991)\citenamefont{Meir, Wingreen, and
  Lee}}]{Transport_QuantumDot}
\bibinfo{author}{\bibfnamefont{Y.}~\bibnamefont{Meir}},
  \bibinfo{author}{\bibfnamefont{N.~S.} \bibnamefont{Wingreen}},
  \bibnamefont{and} \bibinfo{author}{\bibfnamefont{P.~A.} \bibnamefont{Lee}},
  \bibinfo{journal}{Phys. Rev. Lett.} \textbf{\bibinfo{volume}{66}},
  \bibinfo{pages}{3048} (\bibinfo{year}{1991}).

\bibitem[{\citenamefont{Beenakker}(1991)}]{CoulombBlockade_Beenakker}
\bibinfo{author}{\bibfnamefont{C.~W.~J.} \bibnamefont{Beenakker}},
  \bibinfo{journal}{Phys. Rev. B} \textbf{\bibinfo{volume}{44}},
  \bibinfo{pages}{1646} (\bibinfo{year}{1991}).

\bibitem[{\citenamefont{Qi et~al.}(2008)\citenamefont{Qi, Hughes, and
  Zhang}}]{ShouChengZhang_charge_pump}
\bibinfo{author}{\bibfnamefont{X.-L.} \bibnamefont{Qi}},
  \bibinfo{author}{\bibfnamefont{T.~L.} \bibnamefont{Hughes}},
  \bibnamefont{and} \bibinfo{author}{\bibfnamefont{S.-C.} \bibnamefont{Zhang}},
  \bibinfo{journal}{Nature Physics} \textbf{\bibinfo{volume}{4}},
  \bibinfo{pages}{273} (\bibinfo{year}{2008}).

\bibitem[{\citenamefont{Yasui et~al.}(2012)\citenamefont{Yasui, Itakura, and
  Nitta}}]{NuclearPhysicsB_DiracFermion}
\bibinfo{author}{\bibfnamefont{S.}~\bibnamefont{Yasui}},
  \bibinfo{author}{\bibfnamefont{K.}~\bibnamefont{Itakura}}, \bibnamefont{and}
  \bibinfo{author}{\bibfnamefont{M.}~\bibnamefont{Nitta}},
  \bibinfo{journal}{Nuclear Physics B} \textbf{\bibinfo{volume}{859}},
  \bibinfo{pages}{261} (\bibinfo{year}{2012}).

\bibitem[{\citenamefont{Wu et~al.}(2020{\natexlab{b}})\citenamefont{Wu, Liu,
  Liu, Jiang, and Xie}}]{YJWu_NSR}
\bibinfo{author}{\bibfnamefont{Y.}~\bibnamefont{Wu}},
  \bibinfo{author}{\bibfnamefont{H.}~\bibnamefont{Liu}},
  \bibinfo{author}{\bibfnamefont{J.}~\bibnamefont{Liu}},
  \bibinfo{author}{\bibfnamefont{H.}~\bibnamefont{Jiang}}, \bibnamefont{and}
  \bibinfo{author}{\bibfnamefont{X.~C.} \bibnamefont{Xie}},
  \bibinfo{journal}{National Science Review} \textbf{\bibinfo{volume}{7}},
  \bibinfo{pages}{572} (\bibinfo{year}{2020}{\natexlab{b}}).

\end{thebibliography}

\end{document}